\documentclass[twocolumn, times]{aastex631}

\accepted{\today}
\submitjournal{ApJ}

\shorttitle{TW Hya Shadows}
\shortauthors{Teague et al.}
\graphicspath{{./}{figures/}}
\begin{document}

\title{Gas and Dust Shadows in the TW~Hydrae Disk}

\correspondingauthor{Richard Teague}
\email{richard.d.teague@cfa.harvard.edu}

\author[0000-0003-1534-5186]{Richard Teague}
\affil{Center for Astrophysics \textbar\ Harvard \& Smithsonian, 60 Garden Street, Cambridge, MA 02138, USA}

\author[0000-0001-7258-770X]{Jaehan Bae}
\affiliation{Department of Astronomy, University of Florida, Gainesville, FL 32611, USA}

\author[0000-0002-7695-7605]{Myriam Benisty}
\affiliation{Univ. Grenoble Alpes, CNRS, IPAG, F-38000 Grenoble, France}

\author[0000-0003-2253-2270]{Sean~M.~Andrews}
\affiliation{Center for Astrophysics \textbar\ Harvard \& Smithsonian, 60 Garden Street, Cambridge, MA 02138, USA}

\author[0000-0003-4689-2684]{Stefano Facchini}
\affiliation{Universit\`a degli Studi di Milano, Via Celoria 16, 20133 Milano, Italy}

\author[0000-0001-6947-6072]{Jane Huang}
\altaffiliation{NASA Hubble Fellowship Program Sagan Fellow}
\affiliation{Department of Astronomy, University of Michigan, 323 West Hall, 1085 S. University Avenue, Ann Arbor, MI 48109, United States of America}

\author[0000-0003-1526-7587]{David Wilner}
\affiliation{Center for Astrophysics \textbar\ Harvard \& Smithsonian, 60 Garden Street, Cambridge, MA 02138, USA}

\begin{abstract}
We present new observations of CO $J = 2-1$ emission from the protoplanetary disk around TW~Hya. Emission is detected out to $240~{\rm au}$ ($4\arcsec$) and found to exhibit azimuthal variations up to 20\% beyond 180~au (3\arcsec{}), with the west side of the disk brighter than the east. This asymmetry is interpreted as tracing the shadow previously seen in scattered light. A reanalysis of the multi-epoch observations of the dust shadow in scattered light from \citet{Debes_ea_2017} suggests that an oscillatory motion would provide a better model of the temporal evolution of the dust shadow rather than orbital motion. Both models predict an angular offset between the dust shadow and the gas shadow of up to ${\sim}~100\degr$. We attribute this offset to the finite rate at which dust grains and gas molecules can exchange heat, dominated by the collisional rate between gas molecules and dust grains, $t_{\rm coll}$. The angular offsets derived are equivalent to collisional timescales that range from the near instantaneous up to $t_{\rm coll} \sim 10~{\rm years}$, depending on whether a straight or curved dust shadow, as suggested by HST observations reported by \citet{Debes_ea_2017}, is adopted. The inferred range of $t_{\rm coll}$ are consistent with those predictions based on representative gas densities, temperatures, gas-to-dust ratios and grain sizes. These results represent the first time empirical constraints can be placed on $t_{\rm coll}$.
\end{abstract}

\section{Introduction}
\label{sec:introduction}

TW~Hya is one of the most well studied protoplanetary disks, owing primarily to its unique combination of proximity to Earth \citep[$d = 60.1$~pc;][]{Gaia_ea_2018} and near face-on orientation \citep[$i \approx 5\fdg8$;][]{Teague_ea_2019a}. As such, it is often used as a young analogue of the Solar System, guiding our understanding of how planetary systems like our own may have formed.

For TW~Hya, the radial distribution of dust grains has been exquisitely mapped, both at (sub-)~mm wavelengths, tracing the thermal emission of the mm-sized grains at the disk midplane \citep{Andrews_ea_2016, Huang_ea_2018, Macias_ea_2021}, and in the optical and NIR, tracing the scattering from sub-\micron{} grains high in the atmosphere of the disk \citep{Krist_ea_2000, Weinberger_ea_2002, Apai_ea_2004, Roberge_ea_2005, Debes_ea_2013, Debes_ea_2016, Akiyama_ea_2015, Rapson_ea_2015, vanBoekel_ea_2017}. These have revealed a striking level of substructure -- concentric gaps and rings on au scales punctuate an otherwise smooth background distribution across the entire radial extent of the disk -- both along the disk midplane as well as in the atmosphere.

In concert, sub-mm interferometers have facilitated a comprehensive study of the gas component of the disk. Physical properties such as the gas temperature structure \citep{Calahan_ea_2021}, column density \citep{Gorti_ea_2011, Bergin_ea_2013, Schwarz_ea_2016, Zhang_ea_2017} and radial extent \citep{Huang_ea_2018} have all been probed through a combination of empirical analyses and complex thermochemical modeling. Similarly, spectral surveys have revealed a stunning level of chemical complexity, yielding unique insights into the ionization structure of the disk \citep{Cleeves_ea_2015}, the location of potentially planet-forming snowlines \citep{Qi_ea_2013}, and the astrochemical heritage of Solar System bodies \citep{Walsh_ea_2016, Loomis_ea_2018, Canta_ea_2021}.

With the data available for TW~Hya continuing to grow, a new frontier has opened up: temporal variability. Using archival HST data, \citet{Debes_ea_2017} demonstrated that a previously detected shadow in the outer disk, $r \gtrsim 50$~au \citep{Roberge_ea_2005}, appears to move across the face of the disk at a rate of $22\fdg7~{\rm yr^{-1}}$ in an anti-clockwise direction (increasing PA). Curiously, this rotation appears to be counter to the disk rotation direction inferred through the winding direction of spirals, which are assumed to be trailing, reported in \citet{Teague_ea_2019a}. Recent simulations presented by \citet{Nealon_ea_2019, Nealon_ea_2020} demonstrated that a precessing inner disk could also be a plausible scenario to explain the counter rotation.

In this paper we present new observations of CO $J = 2-1$ emission from TW~Hya at a $0\farcs15$ angular resolution which we describe in Section~\ref{sec:observations}. In Section~\ref{sec:ALMA}, we show that there are azimuthal variations, consistent with the previously detected dust shadows. Motivated by the location of the gas shadows, a reanalysis of the reported movement of the HST shadows in Section~\ref{sec:HST} suggests an oscillatory motion is also a plausible model for the temporal evolution of the dust shadow. With this model in hand, Section~\ref{sec:offsets} uses the relative offset between the gas and dust shadows to place unique constraints on the thermal coupling between gas and dust in the outer disk. These new findings are discussed in Section~\ref{sec:discussion} and summarized in Section~\ref{sec:summary}.

\section{Observations}
\label{sec:observations}

Observations were acquired as part of project 2018.A.00021.S (PI: Teague), designed for high spatial and spectral resolution observation of $^{12}$CO $J = 2-1$, CN $N = 2-1$ and CS $J = 5-4$. The correlator was set up to cover the $^{12}$CO and CS line with a channel spacing of 15.3~kHz (resulting in a 30.5~kHz resolution, equivalent to $40~{\rm m\,s^{-1}}$ and 230.538~GHz), while the CN was observed at a coarser 61~kHz channel spacing (yielding a 122~kHz resolution, equivalent to $160~{\rm m\,s^{-1}}$) in order to have bandwidth sufficiently wide to fully cover the $J = 5/2 - 3/2$ and $J = 3/2 - 1/2$ fine structure groups. A single spectral window in FDM mode centered at 241.5~GHz was used for continuum emission in order to self-calibrate the data. A serendipitous detection of CH$_2$CN in the continuum window was presented in \citet{Canta_ea_2021}. In this paper we focus on the CO emission, leaving analysis of the CS and CN for future work.

\subsection{Calibration}

\begin{figure*}
    \centering
    \includegraphics[width=\textwidth]{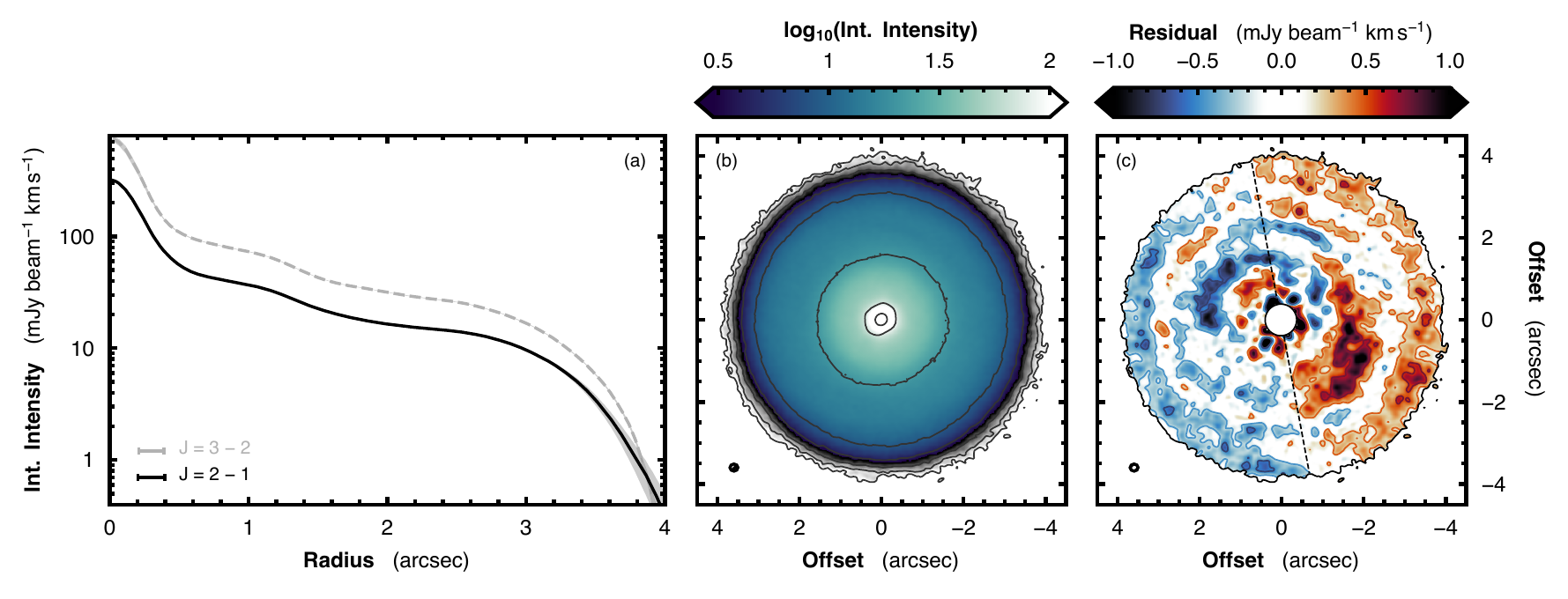}
    \caption{Summary of the CO integrated intensity. (a) The azimuthally averaged radial profile of the zeroth moment map for the $J = 2-1$ emission in black, and the $J = 3 - 2$ emission from \citet{Huang_ea_2018} in dashed gray. The shaded region about each line shows the standard deviation of each annulus and the error bar in the lower left showing the beam FWHM for each data set. (b) The $J = 2-1$ moment map with the beam size shown by the ellipse in the lower left corner. The solid contours show SNR of $5 \times 2^n$ with $n = \{0,\, 1,\, \dots,\, 7\}$. (c) The residuals after subtracting the azimuthally averaged radial profile from the data. The color scaling has been adjusted such that residuals smaller than $1\sigma$ are white. Solid blue and red lines show negative and positive 2 and $5~\sigma$ contours. The dashed line is a guide to emphasize the global east/west asymmetry.}
    \label{fig:moment_maps}
\end{figure*}

Short spacing data (baselines of 15~m to 500~m) were taken 2019 April 4th with two executions including 47.1 minutes of on-source time each. During these executions the quasars J1037-2934 and J1147-3812 were observed for calibration purposes, the former acted as bandpass and flux calibrator and the latter as phase calibrator. Long baseline data (baselines ranging between 15~m and 2.62~km) were taken on 2019 Sep 29th. Only 4 out of the requested 12 executions were observed due to scheduling constraints. Each execution included 50.1 minutes of on source time. During these observations J1037-2934 was observed for bandpass calibration and flux calibration and J1126-3828 for phase calibration. The total on-source time was 4.9 hours.

Initial calibration was performed using the standard pipeline procedure in \texttt{CASA} v5.8.0. The data were then self-calibrated following the procedure used in the DSHARP program \citep{Andrews_ea_2018}. In brief, all spectral windows were used, masking out any lines in each spectral window. These line-free observations were used to derive phase solutions which were then applied to the entire data set. Prior to combining the different executions, each execution was aligned to a common phase center and the continuum fluxes were compared. All executions yielded fluxes that were within 2\% of one another, except for the final long baseline execution which deviated by about 10\%. This execution was rescaled using the \texttt{gaincal} task such that the total flux matched that in the first long baseline execution. The continuum was subtracted using the \texttt{uvcontsub} task.

\subsection{Imaging}

Briggs weighting and a channel spacing of $40~{\rm m\,s^{-1}}$ (the spectral resolution of the data) was chosen for the imaging. A variety of robust parameters were considered, and a value of 0.5 was found to give a good trade off between angular resolution and high quality imaging. This resulted in a synthesized beam of $0\farcs19 \times 0\farcs17$ ($89\degr$). A Keplerian mask was generated,\footnote{\url{github.com/richteague/keplerian\_mask}} making sure that all disk emission was contained within the mask. Following \citet{Teague_ea_2019a}, a stellar mass of $M_{\star} = 0.81~M_{\odot}$ and a viewing geometry described by $i = 5\fdg8$ and ${\rm PA} = 151\degr$ were adopted for this mask. After the imaging, a correction was applied to the image to account for the non-Gaussian synthesized beams due to the combination of several different array configurations \citep[see the discussion in][]{Czekala_ea_2021}, as proposed by \citet{Jorsater_vanMoorsel_1995}. The resulting RMS in a line free channel was measured to be $0.6~{\rm mJy~beam^{-1}}$. Integrating over the Keplerian mask we recover an integrated intensities of $18.31 \pm 0.03~{\rm Jy~km\,s^{-1}}$. This uncertainty does not include a systematic uncertainty of ${\sim}~10\%$ associated with the flux calibration of the data.

The package \texttt{bettermoments}\footnote{\url{github.com/richteague/bettermoments}} was used to calculate the zeroth moment map of the data. This was calculated using the Keplerian \texttt{CLEAN} mask without applying any $\sigma$-clipping. The resulting moment map is shown in Fig.~\ref{fig:moment_maps}b. The analysis of additional moment maps is the focus of a companion paper (Teague et al., in prep.).

\section{Gas Shadows}
\label{sec:ALMA}

No clear azimuthal structure is seen in the zeroth moment map. Thus, to reveal underlying substructure, an azimuthally averaged radial profile was calculated using \texttt{GoFish} \citep{gofish}, adopting the same geometrical properties as used for the \texttt{CLEAN} mask to deproject the data into concentric annuli. To determine the disk center, a Keplerian rotation pattern was fit to the rotation map using \texttt{eddy} \citep{eddy}. This resulted in small offsets relative to the image center of less than a pixel in size: $\Delta {\rm RA} = 20.0 \pm 0.1~{\rm mas}$ and $\Delta {\rm Dec} = 2.0 \pm 0.1~{\rm mas}$. This offset was necessary to include as the centering of the image during the data reduction was performed by a Gaussian fit to the continuum emission. The fitting of a Keplerian rotation pattern allows for a more precise measure of the disk center as both a far larger disk area is used for the fit and the strong azimuthal dependence of the velocity pattern aids in breaking rotational degeneracies of the model. The data was binned into annuli with a width of 1/4 of the beam FWHM (${\approx}~50~{\rm mas}$) to limit the underlying gradient in integrated intensity from biasing the measurement, with the caveat that measurements will be spatially correlated on scales comparable to the beam FWHM. The resulting azimuthally averaged radial profile is shown in Fig.~\ref{fig:moment_maps}a, with a  morphology consistent with that found for the $J=3-2$ line presented by \citet{Huang_ea_2018}.

This radial profile was used to generate an azimuthally symmetric background model which was subtracted from the integrated intensity map shown in Fig~\ref{fig:moment_maps}b. The residuals are shown in Fig.~\ref{fig:moment_maps}c. A clear east/west asymmetry is seen, with the dashed lines marking a rough boundary between the two regimes (the lines trace a PA = 39\degr{}). On top of this large-scale asymmetry, there is much substructure seen within $3\arcsec$, likely associated with the perturbations reported in the gas temperature and dynamics by \citet{Teague_ea_2019a}. To confirm that this asymmetry is unrelated to a misspecified center, the same residuals were calculated for a range of offsets in both the x- and y-direction of $\pm 0.5$~pix (greater than a factor of 10 larger than the statistical uncertainties on the derived source center). In all cases the east/west asymmetry is present.

The global asymmetry is reminiscent of the optical and NIR shadows observed with HST and reported by \citet{Debes_ea_2017}. To make a comparison, we followed the procedure in \citet{Debes_ea_2017} to determine the azimuthal location of the CO shadow. In short, we fit the functional form to each radial bin of the integrated intensity map,
\begin{equation}
    I(r,\, \phi) = \delta I(r) \cdot \cos \big( \phi - \phi_{s,\,{\rm gas}}(r) - \pi \big) + \langle I(r) \rangle,
    \label{eq:I}
\end{equation}
where $\langle I(r) \rangle$ is the azimuthally averaged integrated intensity, $\delta I$(r) is the radially dependent amplitude of the azimuthal variation in integrated intensity and $\phi_{s,\,{\rm gas}}(r)$ is the radially dependent position angle of the shadow (note that unlike \citealt{Debes_ea_2017} we have included a $\pi$ phase offset such that $\phi_{s,\,{\rm gas}}$ describes the azimuthal minimum, not the peak). This form makes the assumption that the shadowing is smooth and affects the whole azimuth, rather than a narrow region with sharp, well defined edges \citep[e.g.,][]{Casassus_ea_2019}. Similarly, it should be noted that $\phi_{s,\,{\rm gas}}$ does not measure the azimuthal minimum of each annulus, but rather than azimuthal location of the sinusoidal model.

The same annuli as those used for the radial profile were adopted for the fitting. For each annulus, the data was averaged over one beam FWHM to minimize the effects of spatial correlations. The package \texttt{emcee} \citep{emcee} was used to sample the posterior distributions of $\{\delta I,\, \phi_{s,\,{\rm gas}}\}$ for each annulus, using 1024 walkers taking 20,000 steps, with the first 10,000 being discarded for burn-in. Uniform priors were adopted for both parameters, with $\delta I$ assumed to be positive and less than $1~{\rm mJy\,beam^{-1}~km\,s^{-1}}$, while $0\degr \le \phi_{s,\,{\rm gas}} < 360\degr$, and no annulus-to-annulus correlation. The results and their associated uncertainties were taken to be the median and 16th to 84th percentile range of the marginalized posterior distributions.

\begin{figure}
    \centering
    \includegraphics[width=\columnwidth]{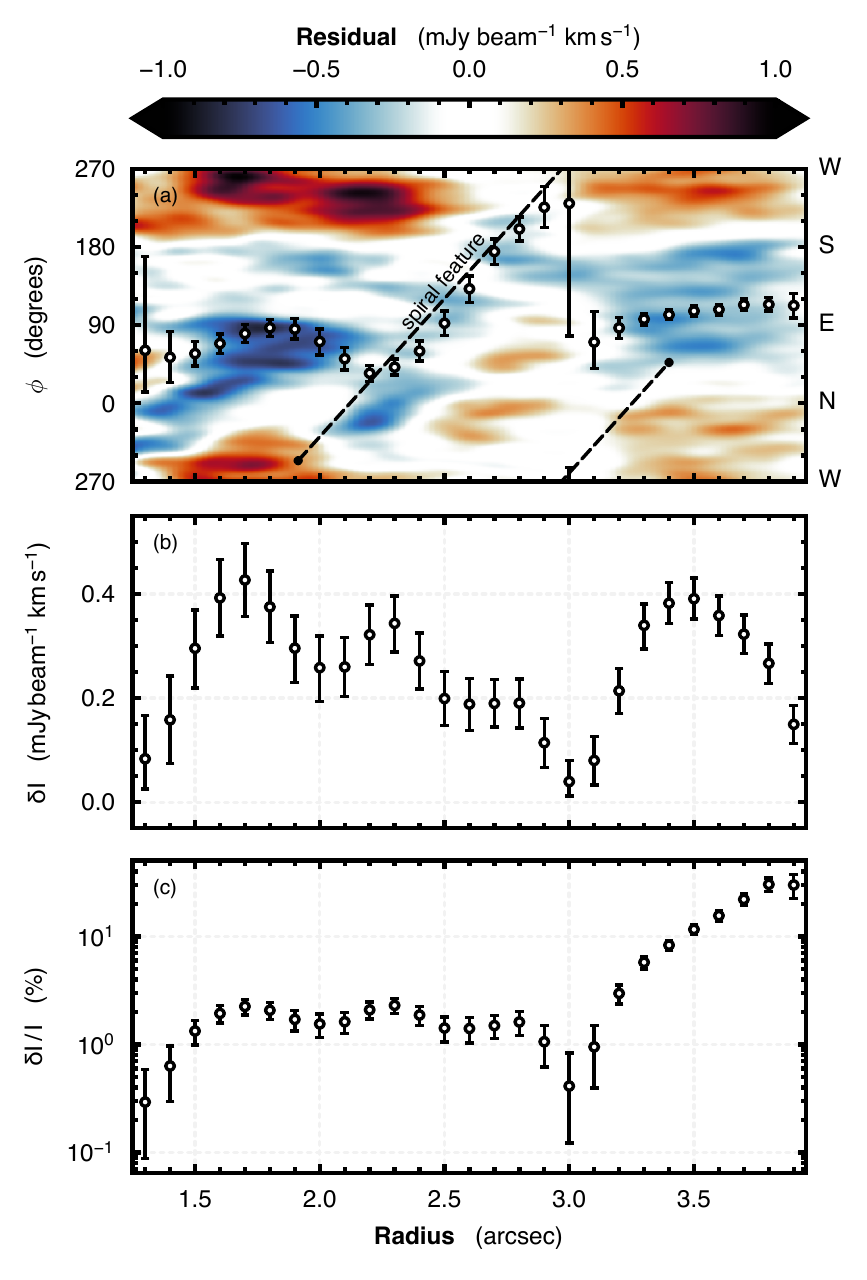}
    \caption{Results of the sub-mm shadow extraction. Panel (a) shows a polar deprojection of the zeroth moment map residuals shown in Fig.~\ref{fig:moment_maps}c. The error bars show the location of $\phi_{s,\,{\rm gas}}$. The outer spiral arm reported by \citet{Teague_ea_2019a} is plotted with a dashed line with a 180\degr{} shift showing that it aligns with the structure within 3\arcsec{}. Panel (b) shows the amplitude of the variation, while panel (c) shows this as a relative fraction of the background average.}
    \label{fig:ALMAshadow}
\end{figure}

The results are shown in Fig.~\ref{fig:ALMAshadow}. Due to the substantial substructure detected in the inner disk \citep[e.g., those discussed in][]{Teague_ea_2019a}, only regions beyond $1\farcs2$ (90~au) are displayed. Inside this radius Eqn.~\ref{eq:I} is a poor description of the azimuthal profile of the integrated intensity. Azimuthal variations of up to ${\approx}~0.4~{\rm mJy\,beam^{-1}~km\,s^{-1}}$ are found across the outer disk, relating to fractional variations of ${\sim}~1\%$ inside of 3\arcsec{}, and increasing to up to ${\sim}~30\%$ outside 3\arcsec{}, as the intensity rapidly drops (see Fig.~\ref{fig:moment_maps}a). The position angle of the shadow appears to have a predominantly easterly direction ($\phi_{s,\,{\rm gas}} \sim 90\degr$), aside from a distinct deviations between 2\arcsec{} and 3\arcsec{}, where $\phi_{s,\,{\rm gas}}$ is likely dominated by the spiral features described in \citet{Teague_ea_2019a}, as can be shown by the black dashed line that shows the spiral peak location (shifted by 180\degr{} to account for the fact $\phi_{s,\,{\rm gas}}$ measures the azimuthal minimum). The reason for the spiral structure only dominating the integrated intensity out to 3\arcsec{} is discussed in Section~\ref{sec:discussion}.

\section{Dust Shadows}
\label{sec:HST}

HST observations have shown that a shadow is present on the surface of the disk at optical and NIR wavelengths \citep{Debes_ea_2017}. Using multi-epoch observations, \citet{Debes_ea_2017} proposed that the shadow moves in an anti-clockwise direction (increasing position angle), at a rate of $22\fdg7~{\rm yr^{-1}}$, equivalent to the orbital frequency of a body at 5.9~au. The authors proposed that a warped inner disk could be capable of casting such a shadow. We developed a model of the temporal evolution of the dust shadow in order to provide a comparison to the gas shadow described in Section~\ref{sec:ALMA}.

\begin{figure*}
    \centering
    \includegraphics[width=\textwidth]{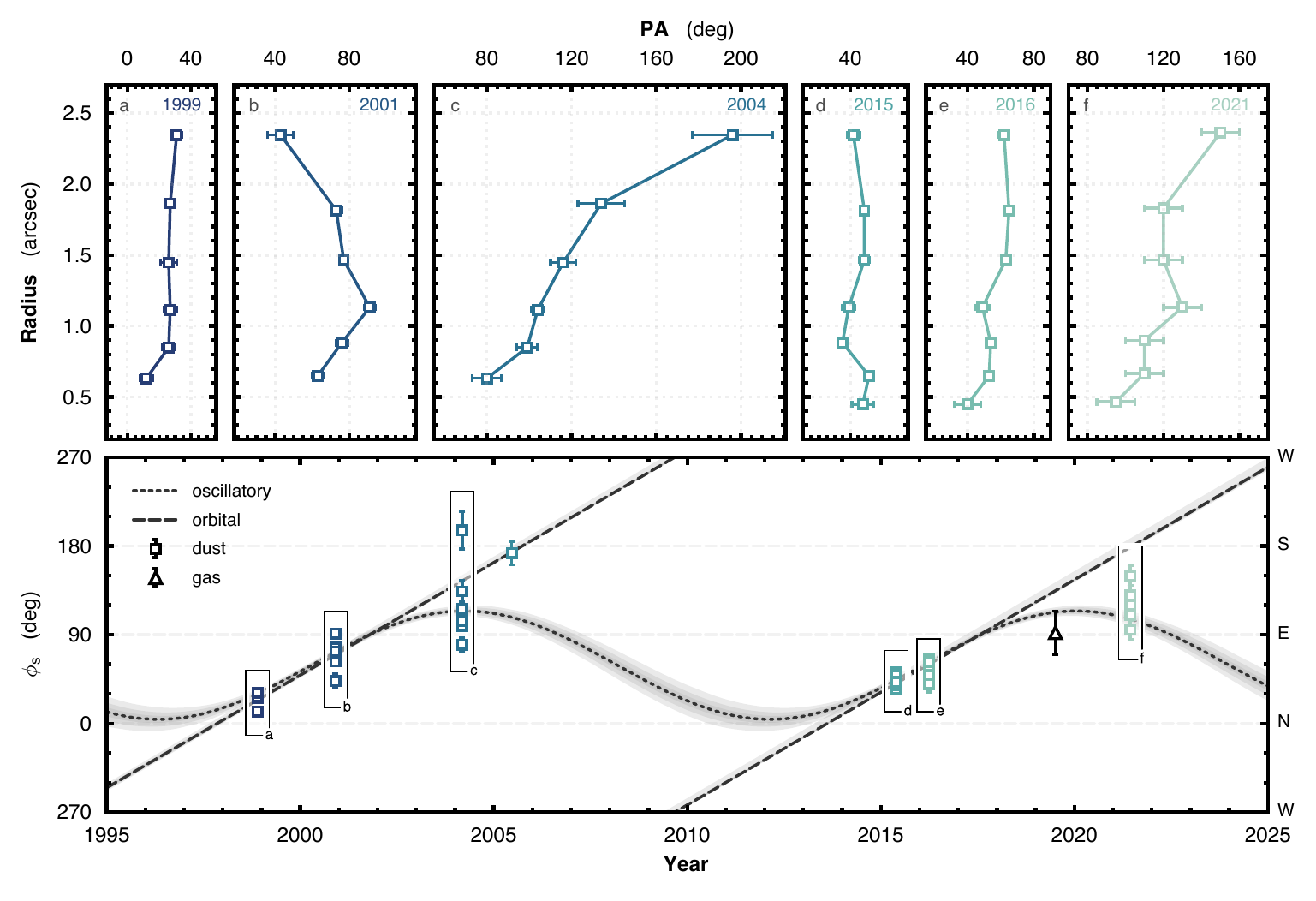}
    \caption{Summary of the temporal evolution of the shadow. The top row shows the six epochs where the shadow has been radially resolved in HST data \citep{Debes_ea_2017}, all sharing the same x-axis scaling. The bottom panel shows the azimuthal location of the shadow as a function of time. Two different models of the time evolution of the shadow are shown: a regular orbit, Eqn.~\ref{eq:phi_orb}, shown by the dashed line, or oscillatory motion, Eqn.~\ref{eq:phi_osc}, shown by the dotted line. The shaded regions about both lines show $\pm 3\sigma$ about the median value of the posterior. Square points show epoch-median shadows observed in the scattered light, with each symbol representing a separate radial bin. The triangle symbol shows the median shadow location beyond 3\arcsec{} observed in the gas with ALMA.}
    \label{fig:HSTshadow_evolution}
\end{figure*}

We used the data presented in \citet{Debes_ea_2017}, supplemented with an additional epoch of data from June 2021 (MJD = 59372) from the on-going HST program \#16228 (J. Debes, private communication). This unpublished data was reduced following the same procedure as described in \citet{Debes_ea_2017}, and all uncertainties were assumed to be $10\degr$ (Debes et al., in prep.). Figure~\ref{fig:HSTshadow_evolution} presents a summary of the data, with the top row showing the radial dependence of the dust shadow for the 6 epochs where it was spatially resolved, while the bottom panel shows the temporal evolution.

\subsection{Curvature}
\label{sec:HST:curvature}

It is clear from the top row of panels in Fig.~\ref{fig:HSTshadow_evolution} that there is a large variation in radial morphology of the dust shadow between epochs: sometimes the shadow appears to have no or minimal radial dependence (1999, 2015 and 2016), while at other times there is a distinct radial structure (2001, 2004 and 2021). As the disk is believed to rotate in a clockwise direction \citep[i.e., decreasing PA;][]{Teague_ea_2019a}, two of these epochs show an apparent `trailing' spiral morphology (2004 and 2021), while the 2001 data shows a distinctly kinked morphology, with the outer regions hinting at a `leading' spiral morphology.

\citet{Kama_ea_2016} have proposed that light travel time effects can produce a curved (in a trailing direction) shadow if the object (called henceforth a `rim', although the exact morphology of the obscuring feature is unknown) that casts the shadow moves an appreciable distance during the time if takes for light to reach the outer disk where the shadow is observed. For a geometrically thin, face on disk, appropriate for TW~Hya, \citet{Kama_ea_2016} showed that the radial form of the shadow is given by
\begin{equation}
    \phi_{s,\,{\rm dust}}(r, t) = \phi_{\rm rim}(t) - \Omega_s \cdot \left(\frac{r - r_{\rm rim}}{c}\right),
    \label{eq:phi_r}
\end{equation}
where $\Omega_s$ is the orbital frequency of the shadow, $c$ the speed of light, and $(r_{\rm rim},\, \phi_{\rm rim}(t))$ describing the cylindrical coordinates of the shadowing rim. Adopting $\Omega_s = 22\fdg7~{\rm yr^{-1}}$, the apparent orbital frequency of the shadow, we find ${\rm d}\phi_{s,\,{\rm dust}} / {\rm d}r \approx 4 \times 10^{-4}~{\rm deg~au^{-1}}$, yielding variations of $\lesssim 0\fdg1$ across the radial extent of the shadow, far too small to account for the curvature observed in the 2004 and 2021 epochs.

Alternatively, ignoring the previously constrained $\Omega_s$ from the overall variation of the shadow with time, we instead assumed that $\Omega_s$ is determined by the spatial morphology of the shadow on an epoch-by-epoch basis. Here, we fixed the $\Omega_s$ to be the Keplerian frequency at the location of a shadowing rim, $\Omega_s = \sqrt{GM_*/r_{\rm rim}^3}$, and solved for $r_{\rm rim}$. Determining $r_{\rm rim}$ for each of the six spatially resolved epochs yielded values of $r_{\rm rim}$ ranging between $r_{\rm rim} = 0.04 \pm 0.01$~au for 2004 to $r_{\rm rim} = 0.8 \pm 0.6$~au for 2015. Such a vast variation in $r_{\rm rim}$ is hard to reconcile with a model of a precessing inner disk \citep[e.g.][]{Facchini_ea_2018, Nealon_ea_2019, Nealon_ea_2020}, and suggests that the inner disk may be highly variable.

It was also argued in \citet{Kama_ea_2016} that the flared scattering surface could make a shadow appear curved, both because of the additional light travel time (the distance the light travels is the spherical polar radius of the disk, rather than the cylindrical radius), and the projection of the shadow on the sky plane. The additional distance due to a flared disk will only increase the light travel time by $\approx 5\%$ for a scattering surface described by $z/r = 0.3$ and is thus unable to account for the observed curvature. To explore the possibility of projection effects creating curvature, we used the Python package \texttt{GoFish} \citep{gofish} to model a range of flared emission surfaces and found that, given the low inclination of TW~Hya, a flared scattering surface could only account for azimuthal variations of $\lesssim 0\fdg1$ across the radial extent of the shadow.

The clear, temporally dependent, structure observed in the scattered light shadow is therefore likely to be reflective of a complex and highly variable morphology for the structure casting the shadow. As a comprehensive model of the inner disk of TW~Hya is not the focus of this paper, a radially constant shadow is assumed for the rest of this analysis. Continued monitoring of the shadow at optical and NIR wavelengths will be essential to characterize the temporal variations and aid in unraveling the inner disk structure.

\subsection{Temporal Evolution}
\label{sec:HST:temporal}

In addition to the change in radial morphology of the shadow, the shadow as a whole is observed to move across the surface of the disk. In order to compare the gas and dust shadows, a model must be used to infer where the dust shadow lies at the time of the ALMA observations.\footnote{As the ALMA observations consist of two sets of observations corresponding to the two different antenna configurations used, the date halfway between both runs is adopted as the time of the ALMA observations: July 1st, 2021.}

First, we verify the angular frequency of the shadow under the assumption of orbital motion by fitting the temporal evolution of the location of the shadow with the form,
\begin{equation}
    \bar{\phi}_{s,\,{\rm dust}}(t) = \big(\bar{\Omega}_s \cdot (t - \delta t)\big) \,\, {\rm mod} \,\, 360\degr
    \label{eq:phi_orb}
\end{equation}
where $\bar{\Omega}_s$ is the angular frequency of the shadow and $\delta t$ is a phase offset. As in Section~\ref{sec:ALMA}, \texttt{emcee} was used to sample the posterior distributions of $\{\bar{\Omega}_s,\, \delta t\}$ using 1024 walkers taking 20,000 steps, with the first 10,000 being discarded for burn-in. The 50th percentile and 16th to 84th percentile range of the posterior distributions were adopted as the best-fit value and uncertainty, respectively. We stress that these uncertainties only represent the statistical uncertainty on these parameters, and do not reflect the systematic uncertainties associated with model selection. The angular frequency of the shadow was found to be $\bar{\Omega}_s = 22\fdg9 \pm 0\fdg1~{\rm yr^{-1}}$, consistent with the angular frequency reported in \citet{Debes_ea_2017}, and shown in the bottom panel of Fig.~\ref{fig:HSTshadow_evolution} as a dashed line. 

Both the 2004 data and the new epoch (2021) of HST data appear to show a shadow that is at a smaller PA than would be predicted by this orbital motion, as marked by the dashed lines in Fig.~\ref{fig:HSTshadow_evolution}. This motivates exploration of an alternative model of the temporal evolution of the shadow. One possibility is oscillatory motion, due perhaps to a differentially precessing inner disk \citep[e.g.,][]{Facchini_ea_2018, Nealon_ea_2019, Nealon_ea_2020}. To check the suitability of this model, we fit the data with the form,
\begin{equation}
    \tilde{\phi}_{s,\,{\rm dust}}(t) = \delta \tilde{\phi}_s \sin\left( 2\pi\,\frac{t - \delta t}{\tau} \right ) + \langle \tilde{\phi}_{s,\,{\rm dust}} \rangle,
    \label{eq:phi_osc}
\end{equation}
where $\langle \tilde{\phi}_{s,\,{\rm dust}} \rangle$ is the average position angle of the shadow, $\delta \tilde{\phi}_{s,\,{\rm dust}}$ is the amplitude of the oscillation, $\tau$ is the period of the oscillation and $\delta t$ is a phase offset. Note that a bar symbol is used to denote the assumption of orbital motion (constant angular velocity of the shadow), as in Eqn.~\ref{eq:phi_orb}, while a tilde represents the oscillatory motion of the shadow, as described by Eqn.~\ref{eq:phi_osc}.

Using the same fitting procedure as before, we found $\delta \tilde{\phi}_{s,\,{\rm dust}} = 55\degr \pm 2\degr$, $\langle \tilde{\phi}_{s,\,{\rm dust}} \rangle = 239\degr \pm 2\degr$ and $\tau = 15.8 \pm 0.1~{\rm yr}$. This fit is shown by the dotted line in the bottom panel of Fig.~\ref{fig:HSTshadow_evolution}. To verify that this fit is not biased by the additional epoch of observations, we fit only those epochs reported in \citet{Debes_ea_2017} and find values of $\delta \tilde{\phi}_{s,\,{\rm dust}} = 57\degr \pm 2\degr$, $\langle \tilde{\phi}_{s,\,{\rm dust}} \rangle = 232\degr \pm 3\degr$ and $\tau = 15.8 \pm 0.1~{\rm yr}$. The similarity in inferred parameters suggests that the original datasets already support the oscillatory model.

Unsurprisingly, both models give a similar period of ${\sim}~15.8~{\rm yr}$ due to the timing of the various observations. However, while the oscillatory motion is unable to describe the 2005 data, potentially owing to the lower signal-to-noise of the data compared to the other epochs (J. Debes, private communication), it appears to give a better fit to both the 2004 and 2021 epochs. More specifically, for 2004 the shadow was observed at $\phi_{s,\,{\rm dust}} = 288\degr \pm 37\degr$ (where the quoted uncertainty describes the uncertainty-weighted standard deviation of sample), while the predicted locations were $\bar{\phi}_{s,\,{\rm dust}} = 319\degr \pm 1\degr$ and $\tilde{\phi}_{s,\,{\rm dust}} = 288\degr \pm 2\degr$, for the orbital and oscillatory motion, respectively, with the oscillatory motion finding a much better agreement. In a similar manner the 2021 epoch data showed the shadow located at $\phi_{s,\,{\rm dust}} = 298\degr \pm 20\degr$, for which the orbital model predicts a location of $\bar{\phi}_{s,\,{\rm dust}} = 350\degr \pm 2\degr$ while the oscillatory model predicts $\tilde{\phi}_{s,\,{\rm dust}} = 281\degr \pm 3\degr$, again resulting in a better fit for the oscillatory motion. While these data are suggestive that oscillatory motion would better describe the temporal evolution of the shadow, additional epochs are required to unambiguously distinguish between these scenarios and, as such, both models are used in the remainder of this work.

\section{Offsets Between Gas and Dust Shadows}
\label{sec:offsets}

Although the mass of a protoplanetary disk is dominated by the gas, it is the dust that determines the temperature. Dust grains absorb the stellar radiation which then transfers the heat to the surrounding gas through collisions between gas molecules and dust grains. When a disk region enters a shadow, there is a deficit of photons for the dust grains to absorb and to scatter resulting in a drop in the dust temperature. The gas may not immediately experience a drop in temperature: gas molecules must first collide with the cooler dust grains to transfer some of their excess energy and cool themselves, described by the thermal accommodation timescale, $t_{\rm th}$. This timescale combines the collisional rate of gas and dust molecules, $t_{\rm coll}$, and the speed at which the collisional energy can be converted to internal thermal energy, and is given by
\begin{equation}
 t_{\rm th} = \frac{c_V}{2 k_B \tilde{\alpha}_T}~t_{\rm coll},
\end{equation}
where $c_V$ is the specific heat capacity of the gas, $k_{\rm B}$ is the Boltzmann constant, $\tilde{\alpha}_{\rm T} \approx 0.5$ and is the thermal accommodation coefficient that characterizes the efficiency of the heat transfer between gas molecules and dust grains \citep{Burke_ea_1983}. The prefactor is sufficiently close to unity that $t_{\rm th} \approx t_{\rm coll}$, is assumed implicitly for the remainder of this paper.

Indeed, it has been shown that the gas-dust collisional timescale in the surface layers of protoplanetary disks may not be infinitely short due to low dust abundance \citep{Facchini_ea_2017, Bae_ea_2021}. If $t_{\rm coll}$ is a reasonable fraction of the local orbital period, $t_{\rm orb}$, then an offset is expected between the shadowed location traced by scattered light and the region where the gas is the coldest as the gas takes time to react to the change in dust temperature.

\begin{figure}
    \includegraphics[]{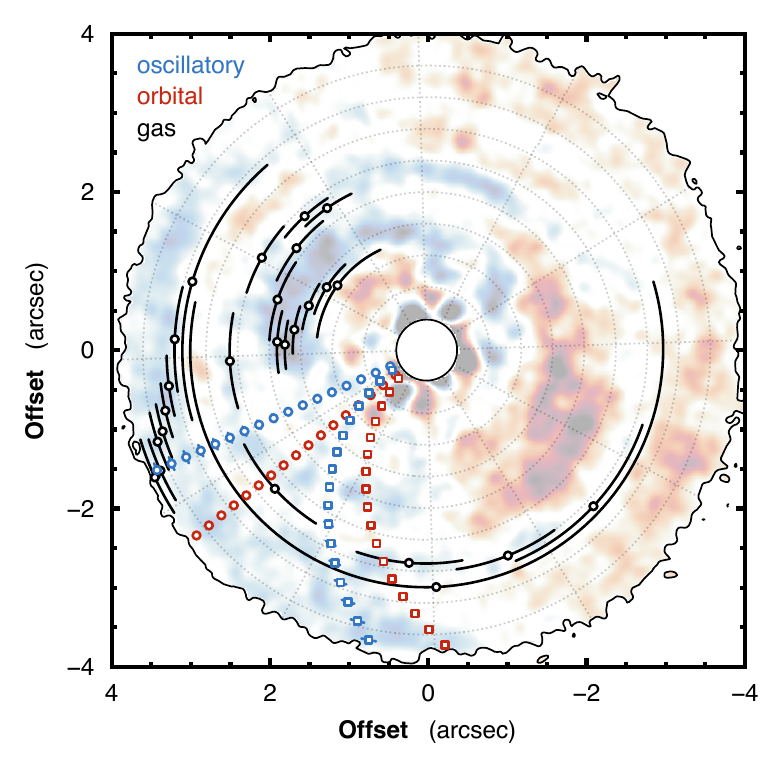}
    \caption{Comparing the gas shadow, black points, with the predicted dust shadow location for both oscillatory and orbital motion, blue and red points, respectively. For each dust shadow model both a straight shadow and curved shadow model are shown with circle and square markers, respectively. The background image is a de-saturated version of Fig.~\ref{fig:moment_maps}c. Note that disk is rotating in a clockwise direction.}
    \label{fig:shadow_comparison}
\end{figure}

\begin{figure*}
    \centering
    \includegraphics[width=\columnwidth]{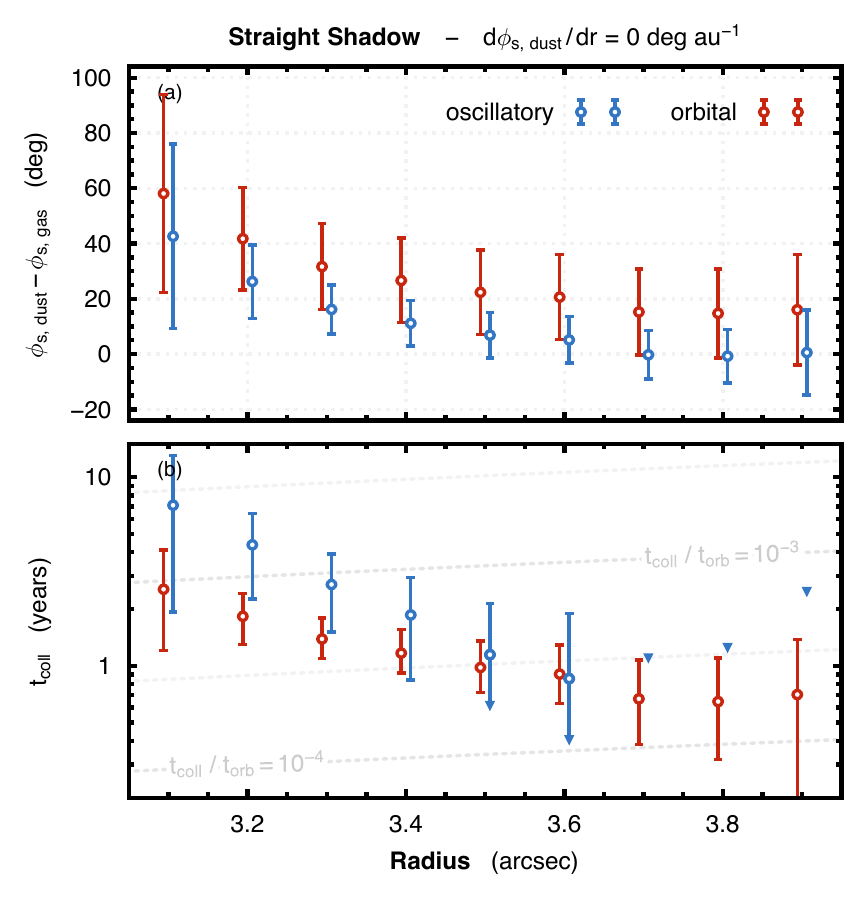}
    \qquad
    \includegraphics[width=\columnwidth]{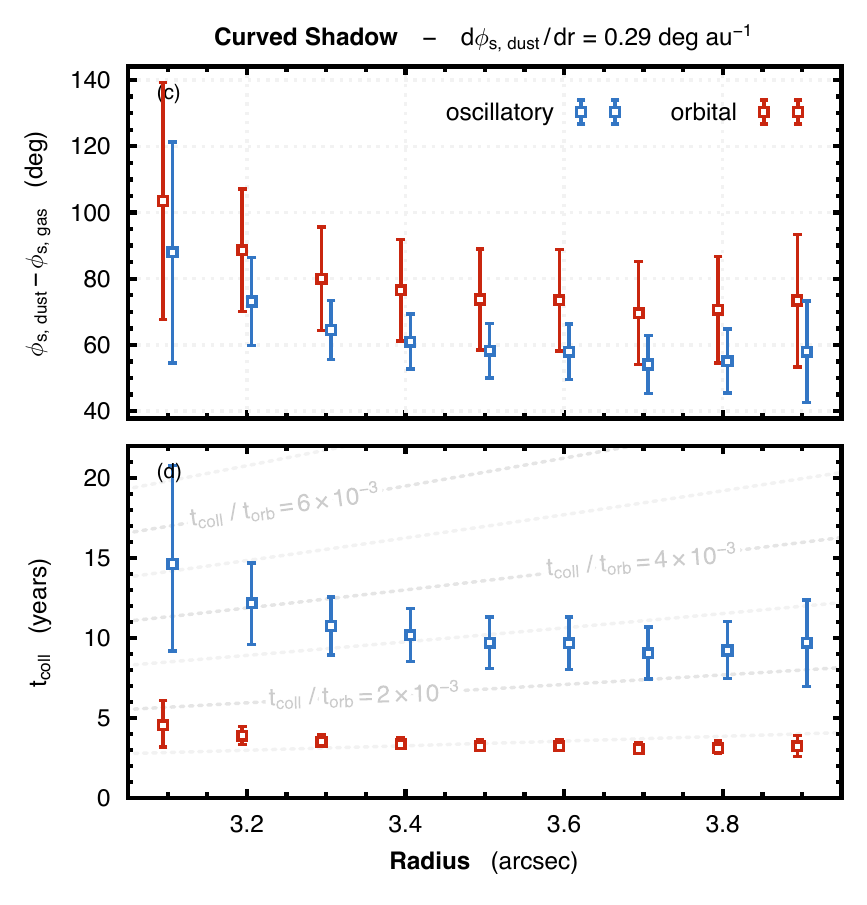}
    \caption{The angular separation between dust and gas shadows and the resulting inferred $t_{\rm coll}$ assuming a straight and curved dust shadow, left and right columns, respectively. In each column, the top panel, (a) and (c), shows the angular separation between the predicted location of the shadow traced by scattered light and traced by CO emission in the outer disk. Red points assume an orbital motion, while blue points assume oscillatory motion. In each column, the bottom panel, (b) and (d), shows the calculated $t_{\rm coll}$ for both temporal models, with dashed lines in the background showing contours of constant $t_{\rm coll} \, / \, t_{\rm orb}$. In both panels the error bars represent $1\sigma$ uncertainties, while arrows represent upper limits. A small radial offset has been applied to both points to avoid overlapping error bars.}
    \label{fig:dphi_tth}
\end{figure*}

Using the models of the temporal evolution of the dust shadows described in the previous section, we calculated the azimuthal location of the shadow at the time of the ALMA observations. As \citet{Debes_ea_2017} does not discuss the presence or absence of shadows beyond $2\farcs4$, we have assumed that the shadow persists to the outer disk where the gas shadow is detected, and modeled the azimuthal offset between dust and gas at different radii assuming both the orbital and oscillatory models. We found an azimuthal offset between the gas and dust shadows of $\sim 60\degr - 0\degr$ between $3\farcs1$ and $3\farcs9$, as shown in Fig.~\ref{fig:shadow_comparison}. To verify that the gas and dust shadows cannot be comoving, we repeated the fitting described in Section~\ref{sec:HST:temporal} including the ALMA data point. There was negligible change in the resulting model parameters confirming that neither the orbital nor the oscillatory model could simultaneously explain the temporal evolution of the gas and dust shadows and that an offset is required, except for the outermost part of the disk. The radially resolved angular separation between shadows is shown in Figs.~\ref{fig:dphi_tth}a and c for a straight and curved shadow, respectively. In each panel both the orbital and oscillatory models of the shadows movement in red and blue, respectively. We propose that this offset could be due to finite gas-dust collisional timescale. 

The angular separation between the shadows in the dust and the gas can therefore be written as
\begin{equation}
    \phi_{s,\, {\rm dust}}(r) - \phi_{s,\, {\rm gas}}(r) = \big(\Omega_s - \Omega_{\rm kep}(r) \big) \cdot t_{\rm coll},
\end{equation}
where $\Omega_s - \Omega_{\rm kep}(r)$ describes the relative angular velocity between the shadow and a parcel of gas in the disk rotating at the Keplerian frequency. Note that while $\Omega_{\rm kep}$ varies radially, $\Omega_s$ is constant as a function of radius as it only depends on the angular velocity of the shadowing rim. At the time of the ALMA observations, both models of the shadows movement predict that the shadow is moving counter-clockwise on the sky (increasing $\phi$), in the opposite direction to the rotation of the disk such that $\partial \phi_{s,\,{\rm dust}} / \partial t$ is positive and $\Omega_{\rm kep}$ is negative. As $\Omega_s$ is considerably larger than $\Omega_{\rm kep}$, any changes in the gas velocity structure, such as substantial sub-Keplerian rotation arising due to a large, negative pressure gradient at the disk edge \citep[e.g.,][]{Dullemond_ea_2020}, would have negligible impact on this result.

Adopting the same stellar mass as used for the Keplerian \texttt{CLEAN} mask in Section~\ref{sec:observations} and the instantaneous angular velocity of the shadow at the time of observations, ${\rm d} \bar{\phi}_{s,\,{\rm dust}} / {\rm d} t = 22 \fdg 6~{\rm yr^{-1}} \pm 0\fdg 1~{\rm yr^{-1}}$ and ${\rm d} \tilde{\phi}_{s,\,{\rm dust}} / {\rm d} t = 5 \fdg 1~{\rm yr^{-1}} \pm 0\fdg 8~{\rm yr^{-1}}$, the angular separation of shadows can be related to $t_{\rm coll}$, shown in Fig.~\ref{fig:dphi_tth}c. Dashed lines in the background of this panel show lines of constant $t_{\rm coll} \, / \, t_{\rm orb}$ ($t_{\rm coll}$ is given by Eqn.~\ref{eqn:tcoll}), where values range between roughly $10^{-4}$ and $10^{-2}$. Beyond $3\farcs5$, the posteriors of $t_{\rm coll}$ are consistent with instantaneous gas-dust collisions for the oscillatory model. We note that it is the much slower angular velocity of the shadow if oscillatory motion is assumed than for the orbital motion that gives rise to a larger $t_{\rm coll}$, despite a smaller angular separation between the gas and dust shadows.

There appears to be a subtle radial decrease in $t_{\rm coll}$ for both models, suggesting potentially a more efficient coupling between gas and grains in the outer disk for the vertical layer probed by CO emission, an idea discussed further in Section~\ref{sec:discussion:tcoll}. An alternative scenario is that some of the radial dependency arises from curvature in the dust shadow (see Section~\ref{sec:HST:curvature}) that was not accounted for in the angular offset between gas and dust shadows. To understand the impact of a curved dust shadow, we repeated the calculation of $t_{\rm coll}$ under the assumption of a curved dust shadow. We adopted a radial morphology given by Eqn.~\ref{eq:phi_r} (i.e., an arithmetic spiral), and linearly interpolated between the best-fit models of the 2016 and 2021 epochs for $\phi_{s,\, {\rm dust}}$ to model the curvature of the shadow during the ALMA observations. This results in a morphology with a radial dependence of ${\rm d}\phi_{s,\, {\rm dust}} / {\rm d}{r} \approx 0\fdg29~{\rm au^{-1}}$, such that average offset between shadows are increased by approximately $70\degr$, as shown in Fig.~\ref{fig:dphi_tth}c. A large angular offset between shadows results in an increase of $t_{\rm coll}$ to ${\sim}~5$~yr for the orbital motion and ${\sim}~15$~yr for the oscillatory motion (see Fig.~\ref{fig:dphi_tth}d). A slight radial dependence of $t_{\rm coll}$ potentially persists, however these results suggest that a curved dust shadow could equally well explain any radial trends.

\section{Discussion}
\label{sec:discussion}

We have shown that CO $J = 2-1$ emission from the disk around TW~Hya shows azimuthal variations on the order of 20\% in the outer disk ($r \gtrsim 3\arcsec$), likely related to a shadow observed in scattered light at smaller radii \citep[$r \lesssim 2\farcs5$;][]{Debes_ea_2017}. Modeling the temporal evolution of the dust shadow suggests an offset between the gas and dust shadow which was interpreted as due to a non-negligible collisional timescale, $t_{\rm coll}$. In this section we discuss the implications of this finding.

\subsection{Collisional Timescale}
\label{sec:discussion:tcoll}

Following \citet{Bae_ea_2021}, the collisional timescale for gas molecules is given by
\begin{eqnarray}
t_{\rm coll} & = & \left( {4 \over 3} \rho_s \right) \left( {\langle a^3 \rangle \over \langle a^2 \rangle} \right) \left( {m_{\rm g} \over \rho_{\rm g} v_{\rm th}} \right) \left( {\rho_{\rm d} \over \rho_{\rm g}} \right)^{-1}, \nonumber \\
 & \simeq & 30.4~{\rm yr} \left( \rho_s \over 3~{\rm g~cm^{-3}} \right) \left( { \langle a^3 \rangle/\langle a^2 \rangle \over 0.1~\mu{\rm m} }\right)  \left( {T_{\rm g} \over 20~{\rm K}} \right)^{-1/2} \nonumber \\
 & & \qquad \times \left( {\rho_{\rm g} \over 10^{-15}~{\rm g~cm}^{-3}} \right)^{-1} \left( { \rho_{\rm d}/\rho_{\rm g} \over 0.01} \right)^{-1},
\label{eqn:tcoll}
\end{eqnarray}
where $\rho_s$ is the bulk density of the dust grains, $\langle a^3 \rangle \, / \, \langle a^2 \rangle$ is the mean dust grain size, $m_g$ is the mean mass of gas molecules, $\rho_g$ is the local gas volume density, $v_{\rm th}$ is the thermal velocity of the gas, and $\rho_d \, / \, \rho_g$ is the local dust-to-gas ratio. We note that $t_{\rm coll}$ therefore scales with the local mean grain size, the local gas-to-dust ratio and inversely with square root of the gas temperature \citep[through the thermal velocity of the gas; see also][]{Young_ea_2004, Facchini_ea_2017}.

With some representative numbers assuming that the line is marginally optically thin, such that it arises from colder regions closer to the midplane, i.e., $\rho_{\rm s} = 3~{\rm g~cm^{-3}}$, $\langle a^3 \rangle \, / \, \langle a^2 \rangle = 0.1 \micron$, $\rho_{\rm g} = 10^{-15}~{\rm g~cm^{-3}}$ \citep[an estimate for one scale height above the midplane assuming a surface density $\Sigma_{\rm g} = 1~{\rm g~cm^{-2}}$ and a gas scale height of 20~au, broadly consistent with the model of][]{Calahan_ea_2021}, $T_{\rm g} = 20$~K, and $\rho_d \, / \, \rho_g = 0.01$, the collisional timescale is $t_{\rm coll} \simeq 30.4~{\rm years}$. This representative value is several times larger than the $t_{\rm coll}$ inferred from the observed shadow offset in TW~Hya (bottom panels of Fig~\ref{fig:dphi_tth} and \ref{fig:dphi_tth}). Such a discrepancy could easily be accounted for with variations of a factor of a few for all the properties described in Eqn.~\ref{eqn:tcoll}. In particular, $\rho_d/\rho_g$ can be one to a few orders of magnitude smaller than 0.01 in surface layers when the settling of dust grains is significant (see e.g., Fig.~2 of \citealt{Bae_ea_2021}), as would be the case for the low levels of turbulence inferred for this disk \citep{Hughes_ea_2011, Teague_ea_2016, Flaherty_ea_2018}.

\subsection{Excitation Conditions}
\label{sec:discussion:excitation}

The shadows described in this work were not observed in previously published high resolution ALMA data of CO emission \citep{Huang_ea_2018, Teague_ea_2019a}. Instead, the previous observations displayed spiral arms that were observed to extend to larger radii (just outside 200~au or $3\farcs4$), before the CO emission was no longer detectable. The difference between these archival observations and the ones presented here are that the previous observations were at slightly higher frequencies, observing the $J = 3-2$ transition with $E_u = 33.2~{\rm K}$, as opposed to the lower energy $E_u = 16.6~K$ of the $J = 2-1$ emission presented here. It is therefore probable that the excitation conditions of the emission influence where and when a shadow can be observed.

In general a lower energy transitions will probe deeper layers in disk, tracing regions closer to the midplane, than higher energy transitions \citep[e.g.,][]{Dartois_ea_2003}. As the vertical settling of dust grains increases with height, such that the local dust-to-gas ratio drops, higher frequency transitions will probe a region of the disk with longer $t_{\rm coll}$. When $t_{\rm coll}$ increases to become a substantial fraction of $t_{\rm orb}$ then a gas parcel will move out of the shadowed region before it has time to collide with the dust to cool down and create a shadow. Therefore, if the $J = 3-2$ emission were sampling a more elevated region in the disk having a longer $t_{\rm coll}$ than the shadow-crossing time this could explain why shadows were not seen, but spirals were (which \citealt{Bae_ea_2021} and \citealt{Muley_ea_2021} argue should get stronger with elevation).

To explore this scenario, we need to consider when a shadow is observable in molecular line emission. The condition for a shadow to form is given by
\begin{equation}
    t_{\rm coll}(r) < {\Delta \phi_{s,\,{\rm dust}}(r) \over |\Omega_s - \Omega_{\rm kep}(r)|},
\label{eqn:shadow_condition}
\end{equation}
where $\Delta \phi_{s,\,{\rm dust}}(r)$ is the radially dependent azimuthal extent of the dust shadow, such that the right hand side is the shadow crossing time. We adopt a radially constant $\Delta \phi_s \simeq 180^\circ$ for the shadow and the maximum angular velocity of the shadow, ${\Omega}_s = 22\fdg9~{\rm yr^{-1}}$ (as any smaller $\Omega_s$ will just increase the limits of $t_{\rm coll}$). As $|\Omega_s| \gg |\Omega_{\rm kep}|$ at these radii, it is irrelevant if the disk is rotating clockwise or anticlockwise, as the relative motion is dominated by the movement of the shadow. We find $t_{\rm coll} \lesssim 8~{\rm years}$ is required for a gas to feel the effects of the shadow. A shadow crossing time of ${\sim}~8$~years sits squarely in the range of $t_{\rm coll}$ inferred for CO $J = 2-1$ considering the different radial and temporal dependencies of the dust shadow. Thus, it is likely the $J = 2-1$ emission probes a vertical layer where $t_{\rm coll}$ is close to this limit, and $J = 3-2$ probes an elevated layer where $t_{\rm coll} > 8$~years due to the decrease in gas and dust density.

This scenario also provides an explanation for why spirals are clearly seen with the $J = 2-1$ emission but are quickly lost beyond $r \sim 3\arcsec$. As the surface density drops towards the outer disk, the optical depth of the line decreases such that the emission probes a progressively less elevated region \citep[which has been directly observed in disks before, e.g.,][]{Teague_ea_2019b, Law_ea_2021}. This would result in the emission tracing progressively smaller $t_{\rm coll}$ values until a threshold is reached (Eqn.~\ref{eqn:shadow_condition}) and shadows can be observed in the gas. A direct test of this scenario would be to search for similar shadows or azimuthal variations in emission which arise from deeper regions of the disk, either the $J = 1-0$ transition of CO, or emission from less abundant isotopologues, and to identify where the shadows first become observable -- although currently no observations of such emission at a sufficient sensitivity or angular resolution exists. 

\section{Summary}
\label{sec:summary}

In this paper we presented new observations of CO $J = 2-1$ emission from the disk around TW~Hya. The integrated intensity exhibits azimuthal variations in the outer disk reaching fractional variations of up to ${\sim}~20\%$ in an east/west direction. These variations are consistent in structure with the azimuthal asymmetries and shadows detected previously in scattered light observations \citep{Roberge_ea_2005, Debes_ea_2017}.

To ascertain any offset in shadows in the dust and gas, observed with HST and ALMA, respectively, the temporal evolution of the shadow was reanalyzed. It was found an oscillatory motion of the dust shadow, rather than orbital motion, would better explain the temporal evolution and readily account for the apparent counter-rotation movement of the shadow over the face of the disk. Both models of the movement of the shadow resulted in a 10\degr{} -- 60\degr{} offset between the dust and gas shadows which was interpreted as due to a non-negligible collisional timescale between the gas molecules and dust grains. Given the uncertainty in the angular velocity of the shadow, this corresponds to a $t_{\rm coll}$ of between 1 and 10 years.

Although the HST data is suggestive of a curved shadow, it is unclear with the current data what could cause this curvature -- both light travel time and projection effects proposed by \citet{Kama_ea_2016} can be ruled out. Adopting a simple linear model of the shadow curvature in the analysis demonstrated that a curved dust shadow would increase the angular separation between dust and gas shadows, resulting in a lengthened $t_{\rm coll}$, ranging between 5 and 20 years.

The inferred $t_{\rm coll}$ timescales are close to the shadow crossing time, suggesting that a small change in local physical conditions would be sufficient for the gas to no longer exhibit a shadow. This is likely the reason the gas shadow is not seen in previous observations of $J = 3-2$ emission as the emission originates from a more elevated region where $t_{\rm coll}$ is larger due to the lower dust densities.

\software{emcee \citep{emcee}, GoFish \citep{gofish}, bettermoments \citep{Teague_Foreman-Mackey_2018}, CASA \citep[v5.8.0;][]{casa}, \texttt{keplerian\_mask.py} \citep{keplerian_mask}}

\begin{acknowledgments}
The authors would like to thank the referee, Takayuki Muto, for their careful and helpful review of the manuscript, John Debes for graciously sharing unpublished HST data from their program \#16228, and enlightening discussion with Kees Dullemond. This paper makes use of the following ALMA data: ADS/JAO.ALMA\#2018.A.00021.S. ALMA is a partnership of ESO (representing its member states), NSF (USA) and NINS (Japan), together with NRC (Canada), MOST and ASIAA (Taiwan), and KASI (Republic of Korea), in cooperation with the Republic of Chile. The Joint ALMA Observatory is operated by ESO, AUI/NRAO and NAOJ. The National Radio Astronomy Observatory is a facility of the National Science Foundation operated under cooperative agreement by Associated Universities, Inc. Based on observations made with the NASA/ESA Hubble Space Telescope, obtained from the Data Archive at the Space Telescope Science Institute, which is operated by the Association of Universities for Research in Astronomy, Inc., under NASA contract NAS5-26555. These observations are associated with program \#16228. This project has received funding from the European Research Council (ERC) under the European Union’s Horizon 2020 research and innovation programme (grant agreement No. 101002188) Support for J. H. was provided by NASA through the NASA Hubble Fellowship grant \#HST-HF2-51460.001-A awarded by the Space Telescope Science Institute, which is operated by the Association of Universities for Research in Astronomy, Inc., for NASA, under contract
NAS5-26555. 
\end{acknowledgments}

\bibliography{main}

\begin{thebibliography}{}
\expandafter\ifx\csname natexlab\endcsname\relax\def\natexlab#1{#1}\fi
\providecommand{\url}[1]{\href{#1}{#1}}
\providecommand{\dodoi}[1]{doi:~\href{http://doi.org/#1}{\nolinkurl{#1}}}
\providecommand{\doeprint}[1]{\href{http://ascl.net/#1}{\nolinkurl{http://ascl.net/#1}}}
\providecommand{\doarXiv}[1]{\href{https://arxiv.org/abs/#1}{\nolinkurl{https://arxiv.org/abs/#1}}}

\bibitem[{{Akiyama} {et~al.}(2015){Akiyama}, {Muto}, {Kusakabe}, {Kataoka},
  {Hashimoto}, {Tsukagoshi}, {Kwon}, {Kudo}, {Kandori}, {Grady}, {Takami},
  {Janson}, {Kuzuhara}, {Henning}, {Sitko}, {Carson}, {Mayama}, {Currie},
  {Thalmann}, {Wisniewski}, {Momose}, {Ohashi}, {Abe}, {Brandner}, {Brandt},
  {Egner}, {Feldt}, {Goto}, {Guyon}, {Hayano}, {Hayashi}, {Hayashi}, {Hodapp},
  {Ishi}, {Iye}, {Knapp}, {Matsuo}, {Mcelwain}, {Miyama}, {Morino},
  {Moro-Martin}, {Nishimura}, {Pyo}, {Serabyn}, {Suenaga}, {Suto}, {Suzuki},
  {Takahashi}, {Takato}, {Terada}, {Tomono}, {Turner}, {Watanabe}, {Yamada},
  {Takami}, {Usuda}, \& {Tamura}}]{Akiyama_ea_2015}
{Akiyama}, E., {Muto}, T., {Kusakabe}, N., {et~al.} 2015, \apjl, 802, L17,
  \dodoi{10.1088/2041-8205/802/2/L17}

\bibitem[{{Andrews} {et~al.}(2016){Andrews}, {Wilner}, {Zhu}, {Birnstiel},
  {Carpenter}, {P{\'e}rez}, {Bai}, {{\"O}berg}, {Hughes}, {Isella}, \&
  {Ricci}}]{Andrews_ea_2016}
{Andrews}, S.~M., {Wilner}, D.~J., {Zhu}, Z., {et~al.} 2016, \apjl, 820, L40,
  \dodoi{10.3847/2041-8205/820/2/L40}

\bibitem[{{Andrews} {et~al.}(2018){Andrews}, {Huang}, {P{\'e}rez}, {Isella},
  {Dullemond}, {Kurtovic}, {Guzm{\'a}n}, {Carpenter}, {Wilner}, {Zhang}, {Zhu},
  {Birnstiel}, {Bai}, {Benisty}, {Hughes}, {{\"O}berg}, \&
  {Ricci}}]{Andrews_ea_2018}
{Andrews}, S.~M., {Huang}, J., {P{\'e}rez}, L.~M., {et~al.} 2018, \apjl, 869,
  L41, \dodoi{10.3847/2041-8213/aaf741}

\bibitem[{{Apai} {et~al.}(2004){Apai}, {Pascucci}, {Brandner}, {Henning},
  {Lenzen}, {Potter}, {Lagrange}, \& {Rousset}}]{Apai_ea_2004}
{Apai}, D., {Pascucci}, I., {Brandner}, W., {et~al.} 2004, \aap, 415, 671,
  \dodoi{10.1051/0004-6361:20034549}

\bibitem[{{Bae} {et~al.}(2021){Bae}, {Teague}, \& {Zhu}}]{Bae_ea_2021}
{Bae}, J., {Teague}, R., \& {Zhu}, Z. 2021, \apj, 912, 56,
  \dodoi{10.3847/1538-4357/abe45e}

\bibitem[{{Bergin} {et~al.}(2013){Bergin}, {Cleeves}, {Gorti}, {Zhang},
  {Blake}, {Green}, {Andrews}, {Evans}, {Henning}, {{\"O}berg}, {Pontoppidan},
  {Qi}, {Salyk}, \& {van Dishoeck}}]{Bergin_ea_2013}
{Bergin}, E.~A., {Cleeves}, L.~I., {Gorti}, U., {et~al.} 2013, \nat, 493, 644,
  \dodoi{10.1038/nature11805}

\bibitem[{{Burke} \& {Hollenbach}(1983)}]{Burke_ea_1983}
{Burke}, J.~R., \& {Hollenbach}, D.~J. 1983, \apj, 265, 223,
  \dodoi{10.1086/160667}

\bibitem[{{Calahan} {et~al.}(2021){Calahan}, {Bergin}, {Zhang}, {Teague},
  {Cleeves}, {Bergner}, {Blake}, {Cazzoletti}, {Guzm{\'a}n}, {Hogerheijde},
  {Huang}, {Kama}, {Loomis}, {{\"O}berg}, {Qi}, {van Dishoeck}, {Terwisscha van
  Scheltinga}, {Walsh}, \& {Wilner}}]{Calahan_ea_2021}
{Calahan}, J.~K., {Bergin}, E., {Zhang}, K., {et~al.} 2021, \apj, 908, 8,
  \dodoi{10.3847/1538-4357/abd255}

\bibitem[{{Canta} {et~al.}(2021){Canta}, {Teague}, {Le Gal}, \&
  {{\"O}berg}}]{Canta_ea_2021}
{Canta}, A., {Teague}, R., {Le Gal}, R., \& {{\"O}berg}, K.~I. 2021, \apj, 922,
  62, \dodoi{10.3847/1538-4357/ac23da}

\bibitem[{{Casassus} {et~al.}(2019){Casassus}, {P{\'e}rez}, {Osses}, \&
  {Marino}}]{Casassus_ea_2019}
{Casassus}, S., {P{\'e}rez}, S., {Osses}, A., \& {Marino}, S. 2019, \mnras,
  486, L58, \dodoi{10.1093/mnrasl/slz059}

\bibitem[{{Cleeves} {et~al.}(2015){Cleeves}, {Bergin}, {Qi}, {Adams}, \&
  {{\"O}berg}}]{Cleeves_ea_2015}
{Cleeves}, L.~I., {Bergin}, E.~A., {Qi}, C., {Adams}, F.~C., \& {{\"O}berg},
  K.~I. 2015, \apj, 799, 204, \dodoi{10.1088/0004-637X/799/2/204}

\bibitem[{{Czekala} {et~al.}(2021){Czekala}, {Loomis}, {Teague}, {Booth},
  {Huang}, {Cataldi}, {Ilee}, {Law}, {Walsh}, {Bosman}, {Guzm{\'a}n}, {Gal},
  {{\"O}berg}, {Yamato}, {Aikawa}, {Andrews}, {Bae}, {Bergin}, {Bergner},
  {Cleeves}, {Kurtovic}, {M{\'e}nard}, {Nomura}, {P{\'e}rez}, {Qi}, {Schwarz},
  {Tsukagoshi}, {Waggoner}, {Wilner}, \& {Zhang}}]{Czekala_ea_2021}
{Czekala}, I., {Loomis}, R.~A., {Teague}, R., {et~al.} 2021, \apjs, 257, 2,
  \dodoi{10.3847/1538-4365/ac1430}

\bibitem[{{Dartois} {et~al.}(2003){Dartois}, {Dutrey}, \&
  {Guilloteau}}]{Dartois_ea_2003}
{Dartois}, E., {Dutrey}, A., \& {Guilloteau}, S. 2003, \aap, 399, 773,
  \dodoi{10.1051/0004-6361:20021638}

\bibitem[{{Debes} {et~al.}(2016){Debes}, {Jang-Condell}, \&
  {Schneider}}]{Debes_ea_2016}
{Debes}, J.~H., {Jang-Condell}, H., \& {Schneider}, G. 2016, \apjl, 819, L1,
  \dodoi{10.3847/2041-8205/819/1/L1}

\bibitem[{{Debes} {et~al.}(2013){Debes}, {Jang-Condell}, {Weinberger},
  {Roberge}, \& {Schneider}}]{Debes_ea_2013}
{Debes}, J.~H., {Jang-Condell}, H., {Weinberger}, A.~J., {Roberge}, A., \&
  {Schneider}, G. 2013, \apj, 771, 45, \dodoi{10.1088/0004-637X/771/1/45}

\bibitem[{{Debes} {et~al.}(2017){Debes}, {Poteet}, {Jang-Condell}, {Gaspar},
  {Hines}, {Kastner}, {Pueyo}, {Rapson}, {Roberge}, {Schneider}, \&
  {Weinberger}}]{Debes_ea_2017}
{Debes}, J.~H., {Poteet}, C.~A., {Jang-Condell}, H., {et~al.} 2017, \apj, 835,
  205, \dodoi{10.3847/1538-4357/835/2/205}

\bibitem[{{Dullemond} {et~al.}(2020){Dullemond}, {Isella}, {Andrews},
  {Skobleva}, \& {Dzyurkevich}}]{Dullemond_ea_2020}
{Dullemond}, C.~P., {Isella}, A., {Andrews}, S.~M., {Skobleva}, I., \&
  {Dzyurkevich}, N. 2020, \aap, 633, A137, \dodoi{10.1051/0004-6361/201936438}

\bibitem[{{Facchini} {et~al.}(2017){Facchini}, {Birnstiel}, {Bruderer}, \& {van
  Dishoeck}}]{Facchini_ea_2017}
{Facchini}, S., {Birnstiel}, T., {Bruderer}, S., \& {van Dishoeck}, E.~F. 2017,
  \aap, 605, A16, \dodoi{10.1051/0004-6361/201630329}

\bibitem[{{Facchini} {et~al.}(2018){Facchini}, {Juh{\'a}sz}, \&
  {Lodato}}]{Facchini_ea_2018}
{Facchini}, S., {Juh{\'a}sz}, A., \& {Lodato}, G. 2018, \mnras, 473, 4459,
  \dodoi{10.1093/mnras/stx2523}

\bibitem[{{Flaherty} {et~al.}(2018){Flaherty}, {Hughes}, {Teague}, {Simon},
  {Andrews}, \& {Wilner}}]{Flaherty_ea_2018}
{Flaherty}, K.~M., {Hughes}, A.~M., {Teague}, R., {et~al.} 2018, \apj, 856,
  117, \dodoi{10.3847/1538-4357/aab615}

\bibitem[{{Foreman-Mackey} {et~al.}(2019){Foreman-Mackey}, {Farr}, {Sinha},
  {Archibald}, {Hogg}, {Sanders}, {Zuntz}, {Williams}, {Nelson}, {de
  Val-Borro}, {Erhardt}, {Pashchenko}, \& {Pla}}]{emcee}
{Foreman-Mackey}, D., {Farr}, W., {Sinha}, M., {et~al.} 2019, The Journal of
  Open Source Software, 4, 1864, \dodoi{10.21105/joss.01864}

\bibitem[{{Gaia Collaboration} {et~al.}(2018){Gaia Collaboration}, {Brown},
  {Vallenari}, {Prusti}, {de Bruijne}, {Babusiaux}, {Bailer-Jones}, {Biermann},
  {Evans}, {Eyer}, {Jansen}, {Jordi}, {Klioner}, {Lammers}, {Lindegren},
  {Luri}, {Mignard}, {Panem}, {Pourbaix}, {Randich}, {Sartoretti}, {Siddiqui},
  {Soubiran}, {van Leeuwen}, {Walton}, {Arenou}, {Bastian}, {Cropper},
  {Drimmel}, {Katz}, {Lattanzi}, {Bakker}, {Cacciari}, {Casta{\~n}eda},
  {Chaoul}, {Cheek}, {De Angeli}, {Fabricius}, {Guerra}, {Holl}, {Masana},
  {Messineo}, {Mowlavi}, {Nienartowicz}, {Panuzzo}, {Portell}, {Riello},
  {Seabroke}, {Tanga}, {Th{\'e}venin}, {Gracia-Abril}, {Comoretto},
  {Garcia-Reinaldos}, {Teyssier}, {Altmann}, {Andrae}, {Audard},
  {Bellas-Velidis}, {Benson}, {Berthier}, {Blomme}, {Burgess}, {Busso},
  {Carry}, {Cellino}, {Clementini}, {Clotet}, {Creevey}, {Davidson}, {De
  Ridder}, {Delchambre}, {Dell'Oro}, {Ducourant},
  {Fern{\'a}ndez-Hern{\'a}ndez}, {Fouesneau}, {Fr{\'e}mat}, {Galluccio},
  {Garc{\'\i}a-Torres}, {Gonz{\'a}lez-N{\'u}{\~n}ez}, {Gonz{\'a}lez-Vidal},
  {Gosset}, {Guy}, {Halbwachs}, {Hambly}, {Harrison}, {Hern{\'a}ndez},
  {Hestroffer}, {Hodgkin}, {Hutton}, {Jasniewicz}, {Jean-Antoine-Piccolo},
  {Jordan}, {Korn}, {Krone-Martins}, {Lanzafame}, {Lebzelter}, {L{\"o}ffler},
  {Manteiga}, {Marrese}, {Mart{\'\i}n-Fleitas}, {Moitinho}, {Mora}, {Muinonen},
  {Osinde}, {Pancino}, {Pauwels}, {Petit}, {Recio-Blanco}, {Richards},
  {Rimoldini}, {Robin}, {Sarro}, {Siopis}, {Smith}, {Sozzetti}, {S{\"u}veges},
  {Torra}, {van Reeven}, {Abbas}, {Abreu Aramburu}, {Accart}, {Aerts},
  {Altavilla}, {{\'A}lvarez}, {Alvarez}, {Alves}, {Anderson}, {Andrei},
  {Anglada Varela}, {Antiche}, {Antoja}, {Arcay}, {Astraatmadja}, {Bach},
  {Baker}, {Balaguer-N{\'u}{\~n}ez}, {Balm}, {Barache}, {Barata}, {Barbato},
  {Barblan}, {Barklem}, {Barrado}, {Barros}, {Barstow}, {Bartholom{\'e}
  Mu{\~n}oz}, {Bassilana}, {Becciani}, {Bellazzini}, {Berihuete}, {Bertone},
  {Bianchi}, {Bienaym{\'e}}, {Blanco-Cuaresma}, {Boch}, {Boeche}, {Bombrun},
  {Borrachero}, {Bossini}, {Bouquillon}, {Bourda}, {Bragaglia}, {Bramante},
  {Breddels}, {Bressan}, {Brouillet}, {Br{\"u}semeister}, {Brugaletta},
  {Bucciarelli}, {Burlacu}, {Busonero}, {Butkevich}, {Buzzi}, {Caffau},
  {Cancelliere}, {Cannizzaro}, {Cantat-Gaudin}, {Carballo}, {Carlucci},
  {Carrasco}, {Casamiquela}, {Castellani}, {Castro-Ginard}, {Charlot},
  {Chemin}, {Chiavassa}, {Cocozza}, {Costigan}, {Cowell}, {Crifo}, {Crosta},
  {Crowley}, {Cuypers}, {Dafonte}, {Damerdji}, {Dapergolas}, {David}, {David},
  {de Laverny}, {De Luise}, {De March}, {de Martino}, {de Souza}, {de Torres},
  {Debosscher}, {del Pozo}, {Delbo}, {Delgado}, {Delgado}, {Di Matteo},
  {Diakite}, {Diener}, {Distefano}, {Dolding}, {Drazinos}, {Dur{\'a}n},
  {Edvardsson}, {Enke}, {Eriksson}, {Esquej}, {Eynard Bontemps}, {Fabre},
  {Fabrizio}, {Faigler}, {Falc{\~a}o}, {Farr{\`a}s Casas}, {Federici},
  {Fedorets}, {Fernique}, {Figueras}, {Filippi}, {Findeisen}, {Fonti},
  {Fraile}, {Fraser}, {Fr{\'e}zouls}, {Gai}, {Galleti}, {Garabato},
  {Garc{\'\i}a-Sedano}, {Garofalo}, {Garralda}, {Gavel}, {Gavras}, {Gerssen},
  {Geyer}, {Giacobbe}, {Gilmore}, {Girona}, {Giuffrida}, {Glass}, {Gomes},
  {Granvik}, {Gueguen}, {Guerrier}, {Guiraud}, {Guti{\'e}rrez-S{\'a}nchez},
  {Haigron}, {Hatzidimitriou}, {Hauser}, {Haywood}, {Heiter}, {Helmi}, {Heu},
  {Hilger}, {Hobbs}, {Hofmann}, {Holland}, {Huckle}, {Hypki}, {Icardi},
  {Jan{\ss}en}, {Jevardat de Fombelle}, {Jonker}, {Juh{\'a}sz}, {Julbe},
  {Karampelas}, {Kewley}, {Klar}, {Kochoska}, {Kohley}, {Kolenberg},
  {Kontizas}, {Kontizas}, {Koposov}, {Kordopatis}, {Kostrzewa-Rutkowska},
  {Koubsky}, {Lambert}, {Lanza}, {Lasne}, {Lavigne}, {Le Fustec}, {Le
  Poncin-Lafitte}, {Lebreton}, {Leccia}, {Leclerc}, {Lecoeur-Taibi},
  {Lenhardt}, {Leroux}, {Liao}, {Licata}, {Lindstr{\o}m}, {Lister}, {Livanou},
  {Lobel}, {L{\'o}pez}, {Managau}, {Mann}, {Mantelet}, {Marchal}, {Marchant},
  {Marconi}, {Marinoni}, {Marschalk{\'o}}, {Marshall}, {Martino}, {Marton},
  {Mary}, {Massari}, {Matijevi{\v{c}}}, {Mazeh}, {McMillan}, {Messina},
  {Michalik}, {Millar}, {Molina}, {Molinaro}, {Moln{\'a}r}, {Montegriffo},
  {Mor}, {Morbidelli}, {Morel}, {Morris}, {Mulone}, {Muraveva}, {Musella},
  {Nelemans}, {Nicastro}, {Noval}, {O'Mullane}, {Ord{\'e}novic},
  {Ord{\'o}{\~n}ez-Blanco}, {Osborne}, {Pagani}, {Pagano}, {Pailler},
  {Palacin}, {Palaversa}, {Panahi}, {Pawlak}, {Piersimoni}, {Pineau}, {Plachy},
  {Plum}, {Poggio}, {Poujoulet}, {Pr{\v{s}}a}, {Pulone}, {Racero}, {Ragaini},
  {Rambaux}, {Ramos-Lerate}, {Regibo}, {Reyl{\'e}}, {Riclet}, {Ripepi}, {Riva},
  {Rivard}, {Rixon}, {Roegiers}, {Roelens}, {Romero-G{\'o}mez}, {Rowell},
  {Royer}, {Ruiz-Dern}, {Sadowski}, {Sagrist{\`a} Sell{\'e}s}, {Sahlmann},
  {Salgado}, {Salguero}, {Sanna}, {Santana-Ros}, {Sarasso}, {Savietto},
  {Schultheis}, {Sciacca}, {Segol}, {Segovia}, {S{\'e}gransan}, {Shih},
  {Siltala}, {Silva}, {Smart}, {Smith}, {Solano}, {Solitro}, {Sordo}, {Soria
  Nieto}, {Souchay}, {Spagna}, {Spoto}, {Stampa}, {Steele},
  {Steidelm{\"u}ller}, {Stephenson}, {Stoev}, {Suess}, {Surdej}, {Szabados},
  {Szegedi-Elek}, {Tapiador}, {Taris}, {Tauran}, {Taylor}, {Teixeira},
  {Terrett}, {Teyssandier}, {Thuillot}, {Titarenko}, {Torra Clotet}, {Turon},
  {Ulla}, {Utrilla}, {Uzzi}, {Vaillant}, {Valentini}, {Valette}, {van Elteren},
  {Van Hemelryck}, {van Leeuwen}, {Vaschetto}, {Vecchiato}, {Veljanoski},
  {Viala}, {Vicente}, {Vogt}, {von Essen}, {Voss}, {Votruba}, {Voutsinas},
  {Walmsley}, {Weiler}, {Wertz}, {Wevers}, {Wyrzykowski}, {Yoldas},
  {{\v{Z}}erjal}, {Ziaeepour}, {Zorec}, {Zschocke}, {Zucker}, {Zurbach}, \&
  {Zwitter}}]{Gaia_ea_2018}
{Gaia Collaboration}, {Brown}, A.~G.~A., {Vallenari}, A., {et~al.} 2018, \aap,
  616, A1, \dodoi{10.1051/0004-6361/201833051}

\bibitem[{{Gorti} {et~al.}(2011){Gorti}, {Hollenbach}, {Najita}, \&
  {Pascucci}}]{Gorti_ea_2011}
{Gorti}, U., {Hollenbach}, D., {Najita}, J., \& {Pascucci}, I. 2011, \apj, 735,
  90, \dodoi{10.1088/0004-637X/735/2/90}

\bibitem[{{Huang} {et~al.}(2018){Huang}, {Andrews}, {Cleeves}, {{\"O}berg},
  {Wilner}, {Bai}, {Birnstiel}, {Carpenter}, {Hughes}, {Isella}, {P{\'e}rez},
  {Ricci}, \& {Zhu}}]{Huang_ea_2018}
{Huang}, J., {Andrews}, S.~M., {Cleeves}, L.~I., {et~al.} 2018, \apj, 852, 122,
  \dodoi{10.3847/1538-4357/aaa1e7}

\bibitem[{{Hughes} {et~al.}(2011){Hughes}, {Wilner}, {Andrews}, {Qi}, \&
  {Hogerheijde}}]{Hughes_ea_2011}
{Hughes}, A.~M., {Wilner}, D.~J., {Andrews}, S.~M., {Qi}, C., \& {Hogerheijde},
  M.~R. 2011, \apj, 727, 85, \dodoi{10.1088/0004-637X/727/2/85}

\bibitem[{{Jorsater} \& {van Moorsel}(1995)}]{Jorsater_vanMoorsel_1995}
{Jorsater}, S., \& {van Moorsel}, G.~A. 1995, \aj, 110, 2037,
  \dodoi{10.1086/117668}

\bibitem[{{Kama} {et~al.}(2016){Kama}, {Pinilla}, \& {Heays}}]{Kama_ea_2016}
{Kama}, M., {Pinilla}, P., \& {Heays}, A.~N. 2016, \aap, 593, L20,
  \dodoi{10.1051/0004-6361/201628924}

\bibitem[{{Krist} {et~al.}(2000){Krist}, {Stapelfeldt}, {M{\'e}nard},
  {Padgett}, \& {Burrows}}]{Krist_ea_2000}
{Krist}, J.~E., {Stapelfeldt}, K.~R., {M{\'e}nard}, F., {Padgett}, D.~L., \&
  {Burrows}, C.~J. 2000, \apj, 538, 793, \dodoi{10.1086/309170}

\bibitem[{{Law} {et~al.}(2021){Law}, {Teague}, {Loomis}, {Bae}, {{\"O}berg},
  {Czekala}, {Andrews}, {Aikawa}, {Alarc{\'o}n}, {Bergin}, {Bergner}, {Booth},
  {Bosman}, {Calahan}, {Cataldi}, {Cleeves}, {Furuya}, {Guzm{\'a}n}, {Huang},
  {Ilee}, {Le Gal}, {Liu}, {Long}, {M{\'e}nard}, {Nomura}, {P{\'e}rez}, {Qi},
  {Schwarz}, {Soto}, {Tsukagoshi}, {Yamato}, {van't Hoff}, {Walsh}, {Wilner},
  \& {Zhang}}]{Law_ea_2021}
{Law}, C.~J., {Teague}, R., {Loomis}, R.~A., {et~al.} 2021, \apjs, 257, 4,
  \dodoi{10.3847/1538-4365/ac1439}

\bibitem[{{Loomis} {et~al.}(2018){Loomis}, {Cleeves}, {{\"O}berg}, {Aikawa},
  {Bergner}, {Furuya}, {Guzman}, \& {Walsh}}]{Loomis_ea_2018}
{Loomis}, R.~A., {Cleeves}, L.~I., {{\"O}berg}, K.~I., {et~al.} 2018, \apj,
  859, 131, \dodoi{10.3847/1538-4357/aac169}

\bibitem[{{Mac{\'\i}as} {et~al.}(2021){Mac{\'\i}as}, {Guerra-Alvarado},
  {Carrasco-Gonz{\'a}lez}, {Ribas}, {Espaillat}, {Huang}, \&
  {Andrews}}]{Macias_ea_2021}
{Mac{\'\i}as}, E., {Guerra-Alvarado}, O., {Carrasco-Gonz{\'a}lez}, C., {et~al.}
  2021, \aap, 648, A33, \dodoi{10.1051/0004-6361/202039812}

\bibitem[{{McMullin} {et~al.}(2007){McMullin}, {Waters}, {Schiebel}, {Young},
  \& {Golap}}]{casa}
{McMullin}, J.~P., {Waters}, B., {Schiebel}, D., {Young}, W., \& {Golap}, K.
  2007, in Astronomical Society of the Pacific Conference Series, Vol. 376,
  Astronomical Data Analysis Software and Systems XVI, ed. R.~A. {Shaw},
  F.~{Hill}, \& D.~J. {Bell}, 127

\bibitem[{{Muley} {et~al.}(2021){Muley}, {Dong}, \& {Fung}}]{Muley_ea_2021}
{Muley}, D., {Dong}, R., \& {Fung}, J. 2021, \aj, 162, 129,
  \dodoi{10.3847/1538-3881/ac141f}

\bibitem[{{Nealon} {et~al.}(2019){Nealon}, {Pinte}, {Alexander}, {Mentiplay},
  \& {Dipierro}}]{Nealon_ea_2019}
{Nealon}, R., {Pinte}, C., {Alexander}, R., {Mentiplay}, D., \& {Dipierro}, G.
  2019, \mnras, 484, 4951, \dodoi{10.1093/mnras/stz346}

\bibitem[{{Nealon} {et~al.}(2020){Nealon}, {Price}, \&
  {Pinte}}]{Nealon_ea_2020}
{Nealon}, R., {Price}, D.~J., \& {Pinte}, C. 2020, \mnras, 493, L143,
  \dodoi{10.1093/mnrasl/slaa026}

\bibitem[{{Qi} {et~al.}(2013){Qi}, {{\"O}berg}, {Wilner}, {D'Alessio},
  {Bergin}, {Andrews}, {Blake}, {Hogerheijde}, \& {van Dishoeck}}]{Qi_ea_2013}
{Qi}, C., {{\"O}berg}, K.~I., {Wilner}, D.~J., {et~al.} 2013, Science, 341,
  630, \dodoi{10.1126/science.1239560}

\bibitem[{{Rapson} {et~al.}(2015){Rapson}, {Kastner}, {Millar-Blanchaer}, \&
  {Dong}}]{Rapson_ea_2015}
{Rapson}, V.~A., {Kastner}, J.~H., {Millar-Blanchaer}, M.~A., \& {Dong}, R.
  2015, \apjl, 815, L26, \dodoi{10.1088/2041-8205/815/2/L26}

\bibitem[{{Roberge} {et~al.}(2005){Roberge}, {Weinberger}, \&
  {Malumuth}}]{Roberge_ea_2005}
{Roberge}, A., {Weinberger}, A.~J., \& {Malumuth}, E.~M. 2005, \apj, 622, 1171,
  \dodoi{10.1086/427974}

\bibitem[{{Schwarz} {et~al.}(2016){Schwarz}, {Bergin}, {Cleeves}, {Blake},
  {Zhang}, {{\"O}berg}, {van Dishoeck}, \& {Qi}}]{Schwarz_ea_2016}
{Schwarz}, K.~R., {Bergin}, E.~A., {Cleeves}, L.~I., {et~al.} 2016, \apj, 823,
  91, \dodoi{10.3847/0004-637X/823/2/91}

\bibitem[{Teague(2019{\natexlab{a}})}]{gofish}
Teague, R. 2019{\natexlab{a}}, The Journal of Open Source Software, 4, 1632,
  \dodoi{10.21105/joss.01632}

\bibitem[{Teague(2019{\natexlab{b}})}]{eddy}
---. 2019{\natexlab{b}}, The Journal of Open Source Software, 4, 1220,
  \dodoi{10.21105/joss.01220}

\bibitem[{{Teague}(2020)}]{keplerian_mask}
{Teague}, R. 2020, {richteague/keplerian\_mask: Initial Release}, 1.0, Zenodo,
  Zenodo, \dodoi{10.5281/zenodo.4321137}

\bibitem[{{Teague} {et~al.}(2019{\natexlab{a}}){Teague}, {Bae}, \&
  {Bergin}}]{Teague_ea_2019b}
{Teague}, R., {Bae}, J., \& {Bergin}, E.~A. 2019{\natexlab{a}}, \nat, 574, 378,
  \dodoi{10.1038/s41586-019-1642-0}

\bibitem[{{Teague} {et~al.}(2019{\natexlab{b}}){Teague}, {Bae}, {Huang}, \&
  {Bergin}}]{Teague_ea_2019a}
{Teague}, R., {Bae}, J., {Huang}, J., \& {Bergin}, E.~A. 2019{\natexlab{b}},
  \apjl, 884, L56, \dodoi{10.3847/2041-8213/ab4a83}

\bibitem[{{Teague} \& {Foreman-Mackey}(2018)}]{Teague_Foreman-Mackey_2018}
{Teague}, R., \& {Foreman-Mackey}, D. 2018, Research Notes of the American
  Astronomical Society, 2, 173, \dodoi{10.3847/2515-5172/aae265}

\bibitem[{{Teague} {et~al.}(2016){Teague}, {Guilloteau}, {Semenov}, {Henning},
  {Dutrey}, {Pi{\'e}tu}, {Birnstiel}, {Chapillon}, {Hollenbach}, \&
  {Gorti}}]{Teague_ea_2016}
{Teague}, R., {Guilloteau}, S., {Semenov}, D., {et~al.} 2016, \aap, 592, A49,
  \dodoi{10.1051/0004-6361/201628550}

\bibitem[{{van Boekel} {et~al.}(2017){van Boekel}, {Henning}, {Menu}, {de
  Boer}, {Langlois}, {M{\"u}ller}, {Avenhaus}, {Boccaletti}, {Schmid},
  {Thalmann}, {Benisty}, {Dominik}, {Ginski}, {Girard}, {Gisler}, {Lobo Gomes},
  {Menard}, {Min}, {Pavlov}, {Pohl}, {Quanz}, {Rabou}, {Roelfsema}, {Sauvage},
  {Teague}, {Wildi}, \& {Zurlo}}]{vanBoekel_ea_2017}
{van Boekel}, R., {Henning}, T., {Menu}, J., {et~al.} 2017, \apj, 837, 132,
  \dodoi{10.3847/1538-4357/aa5d68}

\bibitem[{{Walsh} {et~al.}(2016){Walsh}, {Loomis}, {{\"O}berg}, {Kama}, {van 't
  Hoff}, {Millar}, {Aikawa}, {Herbst}, {Widicus Weaver}, \&
  {Nomura}}]{Walsh_ea_2016}
{Walsh}, C., {Loomis}, R.~A., {{\"O}berg}, K.~I., {et~al.} 2016, \apjl, 823,
  L10, \dodoi{10.3847/2041-8205/823/1/L10}

\bibitem[{{Weinberger} {et~al.}(2002){Weinberger}, {Becklin}, {Schneider},
  {Chiang}, {Lowrance}, {Silverstone}, {Zuckerman}, {Hines}, \&
  {Smith}}]{Weinberger_ea_2002}
{Weinberger}, A.~J., {Becklin}, E.~E., {Schneider}, G., {et~al.} 2002, \apj,
  566, 409, \dodoi{10.1086/338076}

\bibitem[{{Young} {et~al.}(2004){Young}, {Lee}, {Evans}, {Goldsmith}, \&
  {Doty}}]{Young_ea_2004}
{Young}, K.~E., {Lee}, J.-E., {Evans}, Neal~J., I., {Goldsmith}, P.~F., \&
  {Doty}, S.~D. 2004, \apj, 614, 252, \dodoi{10.1086/423609}

\bibitem[{{Zhang} {et~al.}(2017){Zhang}, {Bergin}, {Blake}, {Cleeves}, \&
  {Schwarz}}]{Zhang_ea_2017}
{Zhang}, K., {Bergin}, E.~A., {Blake}, G.~A., {Cleeves}, L.~I., \& {Schwarz},
  K.~R. 2017, Nature Astronomy, 1, 0130, \dodoi{10.1038/s41550-017-0130}

\end{thebibliography}
\bibliographystyle{aasjournal}

\end{document}